\documentclass[useAMS,usenatbib]{mn2e}
\usepackage{graphicx}
\newcommand{\etal}{{\it et al.} }

\title[Frying doughnuts: What can the reprocessing of X-rays to IR tell us about the AGN environment?]{
Frying doughnuts: What can the reprocessing of X-rays to IR tell us about the AGN environment?}

\author[B. McKernan, K.E.S. Ford, N.Chang \& C.S. Reynolds]{B. McKernan$^{1,2}$\thanks{E-mail:bmckernan at amnh.org (BMcK)}, K.E.S. Ford$^{1,2}$, N. Chang$^{1}$ \& C.S. Reynolds$^{3}$ \\
$^{1}$Department of Science, Borough of Manhattan Community College, City University of New York, New York, NY 10007\\
$^{2}$ Department of Astrophysics \& Hayden Planetarium, American Museum of Natural History, New York, NY 10024\\
$^{3}$Department of Astronomy, University of Maryland College Park, College 
Park, MD 21242\\}
\begin{document}

\date{Accepted. Received; in original form}

\pagerange{\pageref{firstpage}--\pageref{lastpage}} \pubyear{2008}

\maketitle

\label{firstpage}

\begin{abstract}
Active galactic nuclei (AGN) produce vast 
amounts of high energy radiation deep in their central engines. X-rays either 
escape the AGN or are absorbed and re-emitted mostly as IR. By studying 
the dispersion in the ratio of observed mid-IR luminosity to observed 2-10keV 
X-ray luminosity ($R_{ir/x}$) in AGN we 
can investigate the reprocessing material (possibly a torus or donut of dust) 
in the AGN central engine, independent of model assumptions.

We studied the ratio of observed mid-IR and 2-10keV X-ray luminosities in a 
heterogeneous sample of 245 AGN from the literature. We found 
that when we removed AGN with prominent jets, $\sim 90\%$ of Type I AGN lay 
within a very tight dispersion in luminosity ratio ($1<R_{ir/x}<30$). This 
implies that the AGN central engine is extremely uniform and models 
of the physical AGN environment (e.g. cloud cover, turbulent disk, 
opening angle of absorbing structures such as dusty tori) must span a 
very narrow range of parameters. We also found that the far-IR(100$\micron$) to
 mid-IR (12$\micron$) observed luminosity 
ratio is an effective descriminator between heavily obscured AGN and
 relatively unobscured AGN.

\end{abstract}

\begin{keywords}
galaxies: active --
galaxies: individual -- galaxies: Seyfert -- techniques: spectroscopic
           -- X-rays:  line -- emission: accretion -- disks :galaxies
\end{keywords}

\section{Introduction}
\label{sec:intro}
Active galactic nuclei (AGN) are powered by the 
accretion of matter onto a supermassive ($\sim 10^{[6,9]}M_{\odot}$) black 
hole. X-ray continuum emission from AGN (without jets) is 
believed to originate in the innermost regions of the AGN whereas IR 
continuum emission is believed to originate in dusty, cooler outer regions. 
The IR emission from AGN is due to a combination of reprocessed higher energy 
emission as well as thermal emission 
from star formation. Much if not most of the IR due to reprocessing comes from
 energy dumped in cool dusty material (mostly out of the observers' X-ray 
sightline) by photons with the most energy (X-rays). A comparison of the 
dispersion in 
the IR to X-ray luminosity ratios ($R_{ir/x}$) of a wide variety of AGN should
 therefore allow us to constrain the range of physical properties and geometry
 of the X-ray reprocessing material in AGN, independent of model 
assumptions.

The standard model of AGN explains the wide range of properties of observed 
AGN in terms of viewing angle, since the equatorial flared disk 
expected at the heart of AGN breaks spherical symmetry \citep{b2}. However, 
partially or heavily obscured AGN need not be edge on, since obscuring 
material only needs to block the observer's sightline (see 
e.g. McKernan \& Yaqoob 1998; Risaliti, Elvis \& Nicastro 2002; Miller, 
Turner \& Reeves 2008). Radiation 
reprocessing occurs in cool (possibly obscuring) material distributed around 
the AGN. Independent of assumptions about the distribution of cool material, 
the \emph{dispersion} in $R_{ir/x}$ can constrain the range of possible 
geometries and physical conditions of 
the obscuring, reprocessing material around the central engine. The 
dispersion of $R_{ir/x}$ among AGN also provides information about 
depopulation or selection effects (bias), if there are values of $R_{ir/x}$ 
that are not favoured by AGN. Genuine depopulation in $R_{ir/x}$ space would 
yield model-independent information on allowed 
configurations of the AGN central engine.

The IR and X-ray emission in AGN have been compared in many previous studies 
\citep{b69,b62,b3,b4,b5,b71}. Most recently \citet{b5}
reveal a strong correlation between modelled intrinsic 2-10keV X-ray 
luminosity and near-IR nuclear luminosity, suggestive of a 'leaky' torus of 
dust. \citet{b71} find a strong correlation 
between mid-IR luminosity and the very hard X-ray (14-195keV) observed 
luminosity, which is unaffected by absorption and indicates that the same 
phenomenon underlies all AGN. However, as yet 
there have been no comparisons of the observed hard 2-10keV X-ray luminosity 
with the IRAS band (12-100$\micron$) IR luminosities. The 2-10keV X-rays are 
energetic enough to emerge with relatively mild attentuation through 
Compton-thin 
obscurers with absorption becoming more substantial above $N_{H}\sim 10^{22} 
\rm{cm}^{-2}$. Attenuated X-rays will mostly re-emerge as IR emission. 
Therefore, in tandem with IRAS band IR emission, 2-10keV X-rays are a good 
probe of 
absorption and subsequent reprocessing in the obscuring material around AGN. 
In this paper, for the first time, we studied the ratio of observed 2-10keV 
X-ray emission to IRAS 
band (12-100$\micron$) IR emission for a large, heterogeneous sample of AGN. 
By studying dispersions in the luminosity ratio 
$R_{ir/x}$ of our sample, we hope to obtain information 
about the \emph{dispersion} of physical properties of the AGN environment, 
independent of model assumptions. 

In section~\ref{sec:sed} we briefly discuss X-ray and IR 
luminosity of AGN, including previous studies which compared emission in the 
two bands, and in section~\ref{sec:obs} we discuss our sample of AGN and 
associated caveats. In section~\ref{sec:xir}, we explore the distribution of 
our sample of AGN in $R_{ir/x}$ parameter space and we use AGN classification 
and well-known AGN to constrain the dispersion of AGN in $R_{ir/x}$. In 
section~\ref{sec:blank} we attempt to rule out selection bias as an 
explanation of the dispersion of AGN in $R_{ir/x}$ space. In 
section~\ref{sec:discuss} we discuss our results and establish 
model-independent constraints on the reprocessing material around AGN. 
Section~\ref{sec:conclusions} outlines our conclusions.

\section{Comparing X-ray and IR emission from AGN}
\label{sec:sed}
Most AGN are characterized by an SED that extends at least from the IR band to
 the hard X-ray band, cutting off at $\sim$50-300 keV 
\citep{b94}. The broadband SED of these AGN can therefore always be 
characterized in a naive way by the slope from the IR band to the X-ray band. 
The IR band emission in most Seyfert AGN generally peaks in the IRAS band 
(12-100$\micron$). The X-ray flux from Seyferts is typically strongly absorbed
 at energies $<$2keV (even at relatively low columns of $N_{H} \sim 10^{21} 
\rm{cm}^{-2}$) and fluxes at $>10$keV are often too low to be strongly 
constrained, so a widely accepted 
measure of X-ray emission from AGN is the observed 2-10keV flux.
 Absorption of 2-10keV X-rays becomes more substantial at columns $N_{H} > \sim 10^{22}
\rm{cm}^{-2}$ and so 2-10keV X-rays are a useful 
probe of Compton-thin absorbers along the AGN sightline, as well as 
reprocessing in the AGN environment. By comparing 
the ratio of observed 2-10 keV X-ray to IRAS-band IR emission 
from AGN, we therefore have (1) an approximate measure of steepness over the 
entire SED in most AGN independent of cosmology and (2) sufficient data
 from AGN to build a statistically meaningful sample. 

The observed 2-10 keV X-ray flux from an AGN may not be the 
\emph{intrinsic} X-ray flux of the AGN. With sufficiently large absorbing 
columns the 2-10keV hard X-ray flux can be very strongly attenuated 
(particularly 
when the absorber is Compton-thick, i.e. $N_{H}>1.5 \times 10^{24} \rm{cm}^{-2}$). 
However, it can been 
difficult to separate AGN with weak observed X-ray flux from those with 
genuinely weak intrinsic X-ray flux, without knowledge of the very hard 
X-ray flux ($>14$ keV or so). With the advent of very hard X-ray detectors 
with sufficient effective area (e.g. SaX, Suzaku and the BAT detector on 
Swift), it has only recently become possible to observe AGN unbiased by X-ray 
absorption (see e.g. \citet{b37,b35,b36,b71}). To illustrate the 
effect of absorption on the intrinsic X-ray luminosity, we shall compare the 
observed 2-10keV flux and the interpolated intrinsic 2-10keV flux inferred 
from very hard X-ray measurements for a number of well-known AGN 
(see \S\ref{sec:xir} below). Throughout this paper, we shall 
refer to the \emph{observed} 2-10keV luminosities of AGN, unless otherwise 
indicated. 

Comparisons of X-ray and IR emission in large samples of AGN 
(and normal galaxies) find significant correlations. \citet{b62} found a 
correlations between 60$\micron$ IRAS luminosity and \emph{Einstein} 
(0.5-4.5keV) soft X-ray luminosity. \citet{b69} found correlations between 
IRAS far-IR luminosity (60$\micron$) with the ROSAT 0.1-2.4keV soft X-ray 
luminosity. Both \citet{b62} and \citet{b69} show a clear 
jump towards higher X-ray emission among broad line galaxies compared with 
narrow line galaxies, suggestive of increasing dominance of non-thermal 
emission.  Correlations from such heterogeneous samples of AGN are important 
since they suggest that the same phenomenon underlies all AGN. More recently, 
\citet{b71} found a strong correlation ($L_{ir} \sim L^{1.25}_{X}$) between 
very hard X-ray luminosity (14-195keV) and  mid-IR luminosity in a uniform 
sample of hard X-ray selected AGN. High angular resolution (nuclear) IR 
studies of AGN have also revealed a correlation between mid-IR luminosity (at 
$6\micron$, $10.5 \micron$ and $12.3\micron$ respectively) and the modelled 
intrinsic hard X-ray luminosity (2-10keV) \citep{b3,b4,b5}. These studies 
found no significant difference in 
the average ratio of mid-IR luminosity to modelled intrinsic hard 
X-ray luminosity for AGN assumed 
to be face-on or edge-on, which they suggest implies the dusty torus is 
clumpy rather than continuous, since X-rays must be leaking out from 
the central regions through the torus \citep{b3,b4,b5}.

Distinguishing models of flared disks (likely 
to be continuous) from models of clumpy clouds of gas and dust is important. 
However, high 
resolution IR studies at present will: a) not resolve most AGN, so their 
statistics will be poor, b) generally include contamination from surrounding 
star-formation and narrow line regions anyway \citep{b5}
and c) potentially lose information about 
the AGN-host galaxy and AGN-starburst connection. Such high resolution IR 
studies are also likely to introduce bias in favour of the nearest AGN and 
restrict the AGN sample size. By including all nuclear IR emission, we avoid 
assumptions about the central engine environment and we span the complete 
nuclear activity phenomenon. Furthermore, modelled intrinsic 2-10keV X-ray 
luminosities are a function of individual 
AGN spectral models which can vary dramatically in the literature often for 
the same dataset (see e.g. \citet{b8} and 
references therein). While estimates of observed 2-10keV X-ray luminosity can 
depend on the continuum model chosen (e.g. FeK$\alpha$ line, reflection 
continuum), ultimately the observed luminosity is determined simply by the 
statistical quality of the simplest model fit to the 2-10keV continuum. A 
statistically good fit to the data converges on the observed value of the 
2-10keV luminosity, regardless of whether the underlying model is actually 
correct. However, inferring the actual intrinsic 2-10keV AGN 
luminosity from a model fit can be complicated. For example, some or most of 
the Fe K or reflected emission could originate in cold, neutral material 
outside the AGN central engine, or absorption by ionized gas (warm absorber 
or Fe edges) could be significant.  Therefore, by using observed 2-10keV 
luminosities, we attempt to avoid assumptions about the AGN environment or 
individual model idiosyncracies in estimating intrinsic X-ray luminosity. 
We can more accurately infer the \emph{intrinsic} 2-10keV X-ray luminosity of 
many of the obscured AGN, by including information from very hard X-ray 
studies of AGN (e.g. \citet{b35,b36}).

\section{The Sample and Caveats}
\label{sec:obs}

\begin{table}
\begin{minipage}{90mm}
\caption{The studies from which our AGN sample was derived \label{tab:sample}.
 Listed are study, year, total number of NED objects 
 listed in each paper and number of unique AGN extracted by us from each study,
 which had both IRAS and 2-10keV X-ray flux measurements on NED. Many AGN were 
included in multiple studies. We did not include AGN with only upper limits 
in either the IRAS band or the 2-10keV band.}
\begin{tabular}{@{}lrrr@{}}
\hline
Study & Year & NED objects 
& Unique AGN \\
\hline
McKernan \etal           & 2007 & 15 & 15\\
Horst \etal		 & 2007 & 29 & 10\\
Buchanan \etal  	 & 2006 & 55 & 14\\
Foschini \etal  	 & 2006 & 19 & 7\\
Shinozaki \etal 	 & 2006 & 51 & 15\\
Botte \etal 		 & 2005 & 10 &  4\\
Nagar, Falcke \& Wilson  & 2005 & 250& 36\\
Peterson \etal 		 & 2004 & 37 & 14\\
Lutz \etal 		 & 2004 & 71 & 12\\
Bian \& Zhao		 & 2003 & 41 & 7\\
Sanders \etal		 & 2003 & 664& 23\\
Schade, Boyle \& Letawsky& 2000 & 76 & 3\\
Grupe \etal		 & 1999 & 76 & 4\\
Leighly			 & 1999 & 23 & 5\\
Malizia \etal 		 & 1999 & 36 & 2\\
Turner \etal 		 & 1999 & 36 & 1\\
Bonatto \etal 		 & 1998 & 60 & 4\\
Gonzalez Delgado \etal   & 1997 & 55 & 3\\
Ho, Filippenko \& Sargent& 1997 & 91 & 13\\
Smith \& Done		 & 1996 & 36 & 2\\
Eracleous \& Halpern 	 & 1994 & 108& 7\\
Rush, Malkan \& Spinoglio& 1993 & 893& 8\\
Boroson \& Green 	 & 1992 & 86 & 31\\
\hline
Total			 &	&2802  &245 \\
\hline
\end{tabular}
\end{minipage}
\end{table}

In the spirit of previous surveys such as \citet{b62}, \citet{b69} and 
\citet{b39}, we assembled our sample from a broad range of AGN studies. 
Table~\ref{tab:sample} lists the twenty three studies of a 
total of 2802 AGN from which we assembled 
245 unique AGN with both IRAS data and 2-10keV data in NED 
\footnote{http://nedwww.ipac.caltech.edu/}. Note that, as
 in previous such heterogeneous studies \citep{b39,b62,b69}, our sample is 
\emph{not} intended to be complete. Rather, our intention was to
 draw upon a wide range of AGN studies, incorporating all types of AGN, in 
order to investigate if any trends in $R_{ir/x}$ emerge even with 
heterogeneous groups of data. The point of this study in effect, is to see 
where AGN lie in 
X-ray-IR luminosity ratio ($R_{ir/x}$) space and whether we can 
in principle deduce model-independent results 
from their dispersion. Clumps and scatters in an $R_{ir/x}$ plot can reveal 
valuable information on AGN depopulation and/or AGN selection effects. In 
future work, we will increase our sample size by expanding on the number of 
surveys in Table~\ref{tab:sample}.

For estimates of the AGN X-ray and IR luminosity, we used NED flux 
measurements in each waveband and the NED luminosity distance estimate, which 
uses a WMAP cosmology \citep{b33}. Where there were multiple flux 
measurements, we used the mean flux value in that waveband. 
This will tend to bias our survey against including flaring periods or 
dramatically low states for individual AGN with multiple flux measurements. 
Generally, the error on the flux measurements listed by NED is $\sim 10-20\%$ 
in the X-ray band and $\sim 5-10\%$ in the IR band. 
However, AGN are by nature variable, sometimes extremely so. 
Of the 245 AGN in our study, 128 have multiple measurements of their 
2-10keV X-ray flux. Of these 128 AGN, roughly half (56/128) vary by a factor 
of two or more, only $\sim$1/5th (24/128) vary by a factor of three or more 
and only $\sim 5\%$ (6/128) vary by an order of magnitude or more. Therefore 
we estimate that the 2-10keV flux in $\sim 55\%$ of the AGN in our 
sample varies by less than a factor of two and the flux in $\sim 80\%$ of the 
AGN in our sample varies by less than a factor of three. In spite of several 
well known exceptions, relatively low X-ray variability among AGN in general 
is consistent with previous studies (see e.g. \citet{b65}).

In the IRAS band, 93 of the 245 AGN in our sample have multiple measurements 
of their IRAS band flux. Of these 93 AGN, roughly one quarter (23/93) vary by 
a factor of two or more, only $\sim 10\%$ (8/93) vary by a factor of three or 
more and only 1/93 varies by an order of magnitude or more. Therefore we 
estimate that in the IR band, the flux in $\sim 75\%$ of the AGN in our 
sample varies by less than a factor of two and the flux in $\sim 90\%$ of the 
AGN in our sample varies by less than a factor of three. If these numbers are 
relatively representative of the AGN in our sample, we estimate that the 
majority of AGN luminosity ratios in this study have errors of no more than a factor of $\sim 3$ in 
both the X-ray and IR luminosities. Note that the 
X-ray and IR measurements used in this paper are almost always 
\emph{not simultaneous}. However, in 
spite of intrinsic AGN variability and non-simultaneity of the data in most 
cases, it turns out that our simple approach reveals strong trends in 
the data (see \S\ref{sec:xir} below).

\begin{table}
\begin{minipage}{85mm}
\caption{Breakdown of our sample by redshift and object classification 
\label{tab:z}. AGN with NED classifications involving prominent jets 
(e.g. BL Lacs, BLRGs) have been excluded. The first 
number in columns 2 $\&$ 3 denotes the total number of Group 1 and Group 2 AGN
 in the corresponding $z$ range (see \S\ref{sec:xir} below for further 
discussion of Groups 1 \& 2 and AGN classification). The numbers in brackets 
denote respectively: the number of Seyfert (Sy) AGN; the number of AGN 
classified as Seyfert plus (Sy$+$) something else (e.g. Starburst, 
$\rm{H}_{\rm{II}}$ region, LINER, QSO); the number of non-Seyfert (NS) AGN.}
\begin{tabular}{@{}lll@{}}
\hline
Redshift Range& Group 1 (Sy,Sy$+$,NS) & Group 2 (Sy,Sy$+$,NS)\\
\hline
$<$0.005 	   & 9(3,6,0) & 53(31,9,13) \\
0.005-0.049 & 56(50,6,0)& 53(32,16,5) \\
0.05-0.099   & 16(12,4,0)  & 3(3,0,0)   \\
0.1-0.199   & 18(14,4,0)  & 0   \\
0.2-0.4   & 13(6,6,1)  & 0   \\
\hline
Total			 &112  &109 \\
\hline
\end{tabular}
\end{minipage}
\end{table}

Table~\ref{tab:z} shows the breakdown of our AGN sample into NED 
classification and redshift. AGN classification and our exclusion of AGN with 
prominent jets is discussed in more detail in \S\ref{sec:xir} below. As far as
 AGN redshift is concerned, most 
($\sim 85\%$) of the AGN in our sample are at low redshift ($z<0.1$), 
so the 12$\micron$ IR flux and 2-10 keV band X-ray flux does not include 
large amounts of flux redshifted from higher energies. However, for a small 
fraction ($\sim 15 \%$) of the higher luminosity AGN, the observed 12$\micron$
 and 2-10 keV fluxes may correspond to fluxes in the AGN frame of $\sim 6-12
\micron$ and $\sim 2-15$keV respectively. Generally in these AGN, 
high energy X-ray flux falls off faster than the near IR flux, so in the
 small fraction of very distant luminous AGN we expect $R_{ir/x}$ to be 
slightly larger \emph{on average} than the nearby group 1 AGN (see also Fig.~\ref{fig:deep} and related discussion below).

In this paper, we discuss luminosity ratios. The IR flux is measured in Jy 
(=$3.0\times10^{-11}$ erg $\rm{cm}^{-2} \rm{s}^{-1}$ at 100$\micron$ and 
$2.5\times10^{-10}$ erg $\rm{cm}^{-2} \rm{s}^{-1}$ at 12$\micron$) 
and the X-ray flux is measured in units of $10^{-11}$ erg $\rm{cm}^{-2} 
\rm{s}^{-1}$. The luminosity (and flux) ratios described hereafter are 
unitless.

\section{Relative X-ray and Infra-red luminosities of AGN}
\label{sec:xir}

In this section we will start by presenting our entire AGN sample in 
$R_{ir/x}$ space using observed 2-10keV X-ray and IRAS-band IR measurements. 
We shall then use AGN classifications to begin constraining 
the dispersions of different AGN types in $R_{ir/x}$. We shall also introduce 
well-known AGN 'typical' of their classification to aid interpretation of the 
AGN distribution. We shall then compare the distribution of AGN in $R_{ir/x}$ 
using far-IR (100$\micron$) and mid-IR (12$\micron$) observed luminosity.

\begin{figure}
\begin{center}
\includegraphics[height=3.35in,width=3.35in,angle=-90]{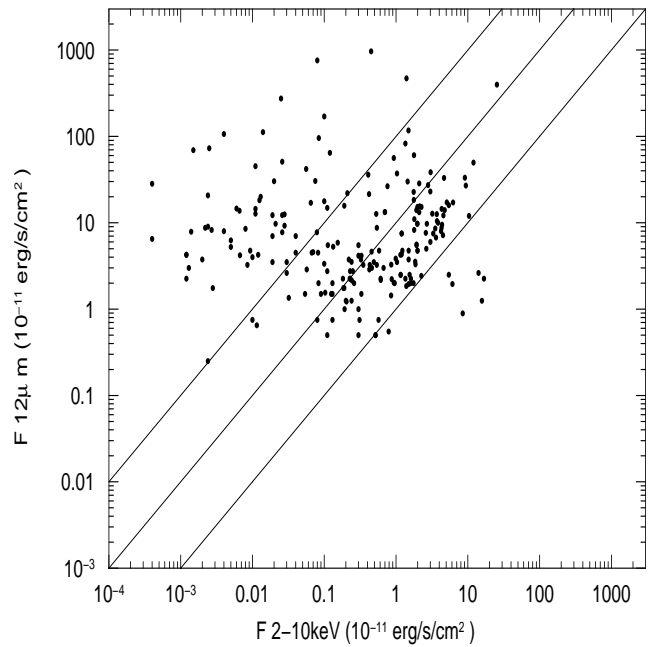}
\caption{Mean 12$\micron$ IR flux (from IRAS/ISO) plotted against mean 
observed 2-10keV X-ray 
flux for 240/245 AGN in our sample (5 AGN had upper limits only at 
12$\micron$). Data (non-simultaneous) are taken from 
NED. Also plotted are lines of constant flux ratio 
(Fir/Fx=1,10,100). }
\label{fig:fir12fx}
\end{center}
\end{figure}

Figure~\ref{fig:fir12fx} shows the observed 2-10keV X-ray flux plotted against 
the 12$\micron$ mid-IR flux for 240/245 AGN in our sample (5 AGN had 12 
$\micron$ upper limits only). Lines of constant flux ratio are plotted 
(Fir/Fx=1,10,100) to guide the eye. If we express the 
flux ratios from Figure~\ref{fig:fir12fx} in terms of luminosity ratios, 
the \emph{ratios} remain constant, but the AGN separate by a factor 
proportional to the distance squared.

\begin{figure}
\begin{center}
\includegraphics[height=3.35in,width=3.35in,angle=-90]{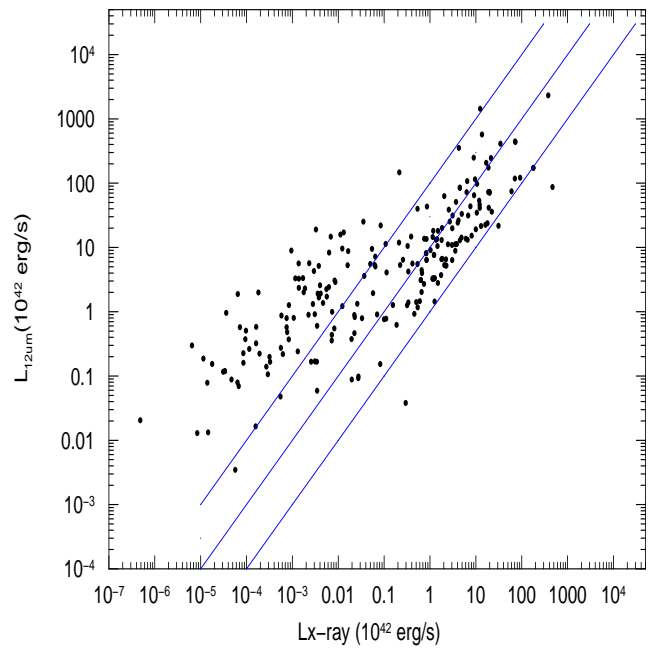}
\caption{Mean 12$\micron$ IR luminosity (from IRAS) plotted against mean 
2-10keV X-ray luminosity for 215/245 AGN in our sample with 
NED classifications that do not include jets. Data (non-simultaneous) are 
taken from NED. Also plotted are lines of $R_{ir/x}$=1,10,100 to guide the 
eye.The three AGN with $R_{ir/x}<1$ are, from bottom to top, Cen A, PG 
1416-129 and PG 0026+129.}
\label{fig:nojets}
\end{center}
\end{figure}

Since jets will complicate our interpretation of the central engine, we 
excluded 20 AGN with NED 
classifications that include prominent jets (e.g. BL Lac, Blazar, LPQ, BLRG). 
Figure~\ref{fig:nojets} shows the mean hard X-ray 2-10keV luminosity plotted 
against the mean mid-IR 12$\micron$ luminosity for 215/245 of the AGN in our 
sample. Lines of constant 
luminosity ratio ($R_{ir/x}$=1,10,100) are indicated
 to guide the eye. From Fig.~\ref{fig:nojets}, the more luminous AGN seem to 
emerge at $L_{x}, L{ir} \sim 10^{42-43}$ ergs $\rm{s}^{-1}$ and mostly lie 
in a band around $R_{ir/x} \sim [1,30]$. Lower luminosity AGN appear to span a
 much wider range of observed luminosities, suggestive
 of both highly absorbed X-ray luminosity and/or a 'noisy' host galaxy 
background in IR or X-rays. The AGN in Fig.~\ref{fig:nojets} span 
$R_{ir/x}$ space in a manner broadly similar to studies carried out using 
soft X-rays \citep{b62,b69}. 

Although we excluded AGN with prominent jets, the remaining AGN may
well contain weak jets, but they are not prominent enough to effect AGN 
classification. AGN with prominent jets accounted for more than half the AGN 
with $L_{x}>3 \times 10^{43}$ ergs $\rm{s}^{-1}$, so if present and prominent,
 jets account for a significant proportion of X-ray luminosity. From 
Fig.~\ref{fig:nojets}, there are only 3/225 AGN with $R_{ir/x}<1$, 
suggesting either that $R_{ir/x}<1$ is atypical for AGN or that there is a 
bias against AGN with $R_{ir/x}<1$ in our sample. It is 
noteworthy that the 3/215 AGN with $R_{ir/x}<1$ in Fig.~\ref{fig:nojets} have 
weak associated jets. They are, in order of 
increasing X-ray luminosity: Cen A, PG 1416-129 and PG 0026+129 respectively 
\citep{b63,b66,b68}. If this interpretation is correct, newly discovered 
AGN with $R_{ir/x}<1$ should have associated jets.

At this point, it is useful to introduce AGN classification 
information (from NED) to aid further interpretation of the dispersion of $R_{ir/x}$ 
values among AGN. Broadly, we observe 'more active' and 'less active' AGN. 
Excluding AGN with jets, the 'more active' AGN population includes Seyfert 1s 
and some QSOs. The 'less active' population generally includes Seyfert 2s, 
LINERs, LLAGN, ULIRGs, some starburst-dominated AGN and some $\rm{H}_{II}$ 
region-like galaxies (see e.g. the definition of 'emission line 
galaxies' for the sample in \citet{b39}). Classifications are of
 course imperfect and so some AGN may overlap between these two groups, e.g. 
Seyfert 1.2-1.9 AGN, as well as cross-classified AGN (e.g. AGN classed both as 
Seyfert 1 and starburst or LINER \& Seyfert 1.5). From Table~\ref{tab:z} above,
of the 221 AGN in our sample without prominent jets, 153/221 AGN are 
classified by NED purely as Seyferts (Sy 1,Sy1.2-1.9, Sy2), 49/221 AGN are 
classified as Sy$+$ or Seyfert plus something else (e.g. starburst, 
$\rm{H}_{\rm{II}}$ region, LINERs) and only 19/221 are non-Seyfert AGN. The 
49/221 AGN classified as Sy$+$ highlight the difficulty of relying too heavily
 on AGN classifications, since the observed properties of many AGN cannot be 
forced into simple classifications. Such cross-classification also 
demonstrates the importance of an approach such as ours; namely a simple 
comparison of observables across a wide range of AGN activity. 

\begin{figure}
\begin{center}
\includegraphics[height=3.35in,width=3.35in,angle=-90]{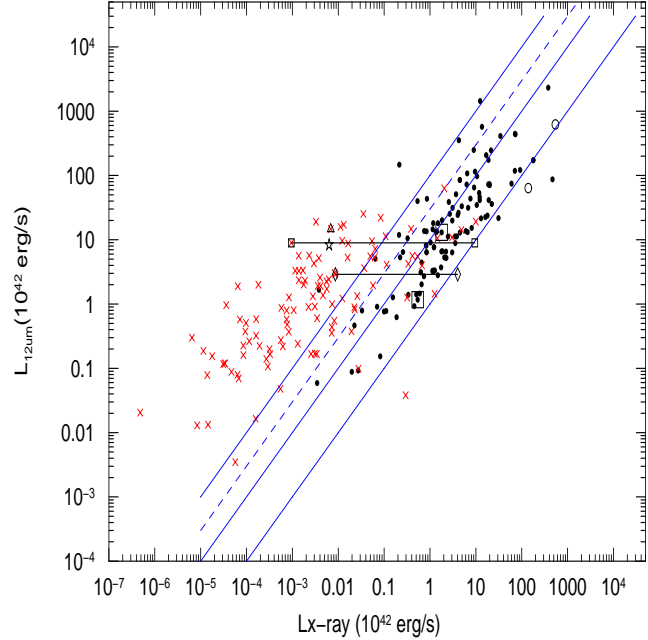}
\caption{As Fig.~\ref{fig:nojets} except we divided our AGN sample into 
the 'more active' Group 1 AGN (black circles) and 'less active' Group 2 AGN 
(red crosses) (see text for details). Dashed line indicates $R_{ir/x}$=30. 
Also highlighted are several 
well-known AGN to aid orientation: Open circles are 3c273 (upper) and 
PKS 2155-304 (lower), examples of a Blazar and BL Lac respectively (Note 
$R<1$ in both cases). Giant open squares (with no lines) are centered on Akn 
564 (upper) and MCG-6-30-15 (lower), examples of a narrow line Seyfert 1 (1.8)
 and Seyfert 1 (1.5) respectively. The open 
triangle is NGC 1068, a Seyfert 2 seen in reflection in X-rays and the open 
star is Arp 220, a starburst galaxy. Open squares with line correspond
 to the observed (left) \& intrinsic (right) X-ray luminosities of NGC 4945 
(\citet{b36}), a Compton-thick AGN seen in transmission. Open diamonds with 
line correspond to the observed (left) \& intrinsic (right) X-ray luminosities
 of the Circinus galaxy (\citet{b38}). Note that the intrinsic X-ray 
luminosities for NGC 4945 \& Circinus were established from very hard X-ray 
measurements (independent of absorption).}
\label{fig:irx12}
\end{center}
\end{figure}

We categorized the remaining 215/245 AGN in our sample based on their 
NED classification. Group 1 includes AGN classified by
 NED as Seyfert 1 and QSOs. Also included in group 1 are AGN which might 
overlap between Groups 1 and 2 (e.g. Seyfert 1.2-1.9 AGN and cross-classified 
AGN). Group 2 AGN includes AGN classified by NED as Seyfert 2, low luminosity 
AGN (LLAGN), LINERs, ULIRGs and starburst dominated AGN. Table~\ref{tab:z} 
shows that most of the AGN in our sample with these latter classifications are
 also classified as Seyfert AGN, again indicating the complexities of AGN 
classifications. Fig.~\ref{fig:irx12}
 shows the two AGN populations highlighted with a dashed line indicating 
$R_{ir/x}=30$. Also indicated are a few very well-known 'archetypal' AGN. We 
included two 
well-known AGN with jets, the Blazar 3C 273 and the BL Lac PKS 2155-304 
(indicated by open circles). These two AGN have some of the smallest observed 
luminosity ratios 
($R_{ir/x} \leq 1$) and are associated with jets close to their sightlines. 
The Seyfert 1 AGN (MCG-6-30-15 and Akn 564, indicated by giant squares) have 
mid-range values of $R_{ir/x}$ for Group 1 AGN.
 The starburst galaxy, Arp 220 (open star) has $R_{ir/x}\sim 1000$, as has the 
Seyfert 2 AGN NGC 1068 (open triangle). The open symbols connected by lines in
 Fig.~\ref{fig:irx12} denote the luminosity ratios for the Compton thick AGN 
NGC 4945 (squares) and Circinus (diamonds) respectively, calculated using both
 the observed (left point) and \emph{interpolated intrinsic} (right point) 
2-10keV luminosity. 

From Fig.~\ref{fig:irx12}, Group 1 AGN (black circles) are the more luminous 
AGN and have a remarkably narrow dispersion in luminosity ratio 
($30<R_{ir/x}<1$) for most (97/107) Group 1 AGN. Of the ten Group 1 AGN with $R_{ir/x}>30$, four 
are cross-classified (i.e. Seyfert 1 plus starburst) and four are heavily 
obscured Seyfert 1s ($N_{H} \sim 10^{23} \rm{cm}^{-2}$). So Group 1 AGN with
$R_{ir/x}>30$ could be accounted for either by the addition of IR luminosity 
from star formation or a drop in observed X-ray luminosity due to a change in 
partial-covering absorption \citep{b91,b92,b93}. So even with our 
simple approach, Fig.~\ref{fig:irx12} indicates that $\sim 90\%$ 
of Group 1 AGN in our sample have $1<R_{ir/x}<30$. By contrast, Group 2 
AGN (red crosses) generally have lower 
X-ray luminosities and have a wider dispersion in $R_{ir/x}$ than Group 1 
AGN. In contrast to the Group 1 AGN, 97/109 group 2 AGN have $R_{ir/x}>30$, 
with 83/109 group 2 AGN having $R_{ir/x}>100$. Of the 12/109 Group 2 AGN with 
$1<R<30$ (i.e. overlapping with the main Group 1 population), we find that six
 have associated radio jets and two are Compton-thin Seyfert 2 AGN. From 
Table~\ref{tab:z}, 19/221 AGN in our sample are non-Seyferts, of which 10 are 
LINERs, 8 LINERs with $H_{\rm{II}}$ regions and 1 QSO. The LINERs lie in 
the bottom left hand corner of the Group 2 distribution ($L_{ir}<0.5 \times 
10^{42}$ erg/s,$L_{x}<10^{39}$ erg/s) and the LINERs with $H_{\rm{II}}$ 
regions span a slightly larger range of the Group 2 distribution ($L_{ir}<5 
\times 10^{42}$ erg/s,$L_{x}<10^{40}$ erg/s).

A key point from Fig.~\ref{fig:irx12} is that the luminosity ratio for NGC 
4945 and Circinus as calculated from the interpolated intrinsic 2-10keV 
luminosity (right point) lies in the main luminosity ratio band of Group 1 
AGN. This is a nice demonstration of the fact that behind the Compton-thick 
obscuration of NGC~4945 and Circinus, lies the same phenomenon as in most 
Group 1 AGN. Without very hard X-ray measurements, it is very difficult to 
estimate intrinsic 2-10keV X-ray luminosities for most Group 2 AGN in our 
sample and certainly any such estimates would be highly model-dependent. 
Nevertheless, the examples of NGC 4945 and Circinus indicate that the 
\emph{intrinsic} luminosity ratios 
for many Group 2 AGN may be similar to the Group 1 luminosity ratio range 
($\sim 30 \leq R_{ir/x} \leq 1$). However, there are many 
model-dependent caveats in estimating the intrinsic X-ray luminosity of an 
AGN, so in \S\ref{sec:discuss} below we will take a different approach to 
\citet{b3,b4,b5} and attempt to derive constraints on AGN using 
\emph{observed} luminosities.

\begin{figure}
\begin{center}
\includegraphics[height=3.35in,width=3.35in,angle=-90]{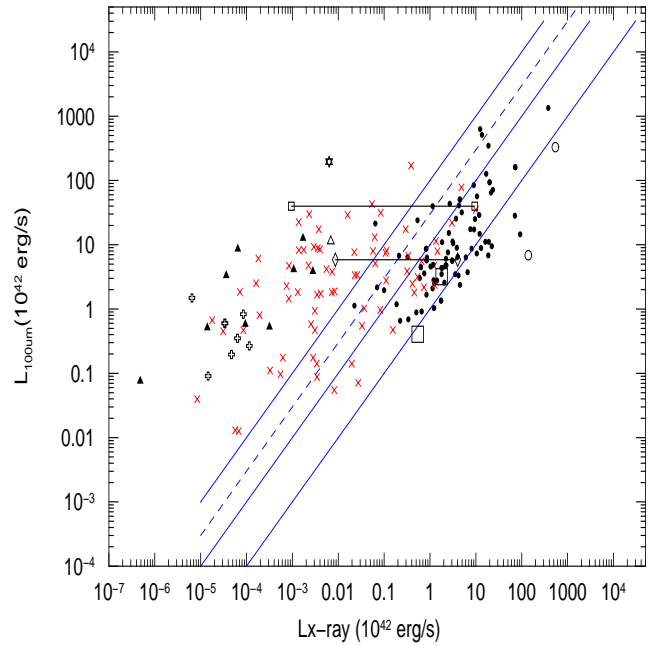}
\caption{As Fig.~\ref{fig:irx12} except we plot mean 
100$\micron$ IR luminosity 
(from IRAS) against mean observed 2-10keV X-ray luminosity for our AGN sample.
 Also highlighted are the non-Seyfert AGN in our sample. Open crosses in the 
lower left-hand corner are LINERs and filled-in triangles denote LINERs plus 
$\rm{H}_{II}$ regions.}
\label{fig:irx100}
\end{center}
\end{figure}

Figure~\ref{fig:irx100} shows the mean far-IR (100$\micron$) luminosity of the
 AGN in our sample versus the mean 2-10keV observed X-ray luminosity. Note 
that several 
AGN had NED flux measurements at $12\micron$ but only upper limits at 
$100\micron$ and vice versa, so 
there is not a perfect one-to-one correspondence between all points on 
Fig.~\ref{fig:irx100} and Fig.~\ref{fig:irx12}. Nevertheless, the 
dispersion in $R_{ir/x}$ for the Group 2 AGN in particular seems to be larger 
in Fig.~\ref{fig:irx100} than in Fig.~\ref{fig:irx12} and the 
various archetypal Group 2 AGN appear more separated in luminosity ratio. For 
example, the starburst galaxy Arp 220 (open star) is much more 
clearly distinguished from the classic Type 2 AGN NGC 1068 (open triangle) 
in Fig.~\ref{fig:irx100} than in Fig~\ref{fig:irx12}. The non-Seyfert AGN, 
e.g. the LINERs and LINERs with $\rm{H}_{II}$ regions, are also highlighted. 
As we should expect from such low X-ray and IR luminosity sources, LINERs 
occupy the lower left-hand corner of the Group 2 AGN dispersion in 
Fig.~\ref{fig:irx100}. The LINERs with associated $\rm{H}_{II}$ regions have 
generally higher IR luminosities than the 'pure' LINERs, but the X-ray 
luminosities remain fairly low.

\begin{figure}
\includegraphics[height=3.35in,width=3.35in,angle=-90]{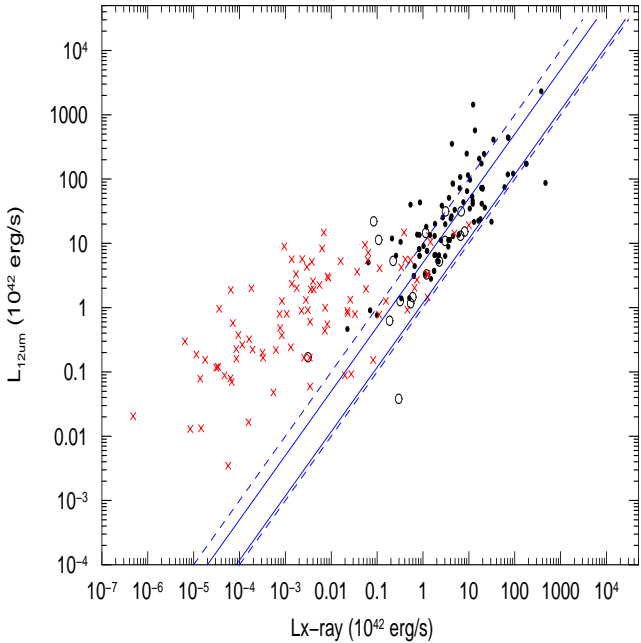}
\caption{As for Fig.~\ref{fig:nojets}, except open circles are AGN from the 
survey of \citet{b5}. Solid lines indicate the range of luminosity ratio 
found by \citet{b5} for group 1 AGN, based on modelled intrinsic 2-10keV X-ray 
luminosity. Dashed lines indicate the range of luminosity ratio found by 
\citet{b5} for group 2 AGN based on modelled intrinsic 2-10keV X-ray 
luminosity.
\label{fig:lutz}}
\end{figure}

Figures~\ref{fig:irx12} and \ref{fig:irx100} both show a very clear 
distinction between the observed luminosity ratios (and their dispersions) for 
group 1 and group 2 AGN. However, based on the modelled intrinsic 
2-10keV X-ray luminosity, \citet{b4,b5} found that the near-IR to X-ray 
luminosity ratios for group 1 and group 2 AGN were very similar. 
Figure~\ref{fig:lutz} is as Fig.~\ref{fig:irx12}, but with open circles 
denoting AGN from the study by \citet{b5} and the ranges of \emph{intrinsic} 
luminosity ratio established by \citet{b5} for group 1 (solid lines) and 
group 2 (dashed lines) AGN. The dispersion in the 
luminosity ratio of our group 1 AGN matches that found by \citet{b5} quite 
well, indicating that observed and intrinsic 2-10keV X-ray luminosities are 
similar in most group 1 AGN. Our group 2 population seems 
to diverge dramatically in luminosity ratio from the range found by 
\citet{b5}, suggesting that the observed and intrinsic 2-10keV luminosities 
in group 2 AGN are indeed dramatically different. However, \citet{b5} 
estimate the intrinsic X-ray luminosity in the group 2 AGN based on different 
model fits in the literature, which (a) can vary for the same AGN dataset and 
which (b) lead to model-dependent constraints 
on AGN structure (e.g. a clumpy torus as suggested by \citet{b5}). By 
contrast, an IR-X-ray luminosity ratio based on the observed 2-10keV X-ray 
luminosity (as in Fig.~\ref{fig:lutz}) will help us avoid model assumptions 
about AGN structure and the AGN environment (see \S\ref{sec:discuss} below).

The observed IR band emission from AGN should consist of some fraction of 
reprocessed radiation from the AGN typically peaking in the mid-IR (e.g. 
\citet{b25}) and some fraction of emission from star formation typically 
peaking in the far-IR (e.g. \citet{b26,b31,b29}) plus intrinsic IR from the 
SED. Therefore, by comparing the 
observed mid- and far-IR luminosities in group 1 and group 2 AGN, we may be 
able to distinguish dominant components of IR in different AGN.

\begin{figure}
\includegraphics[height=3.35in,width=3.35in,angle=-90]{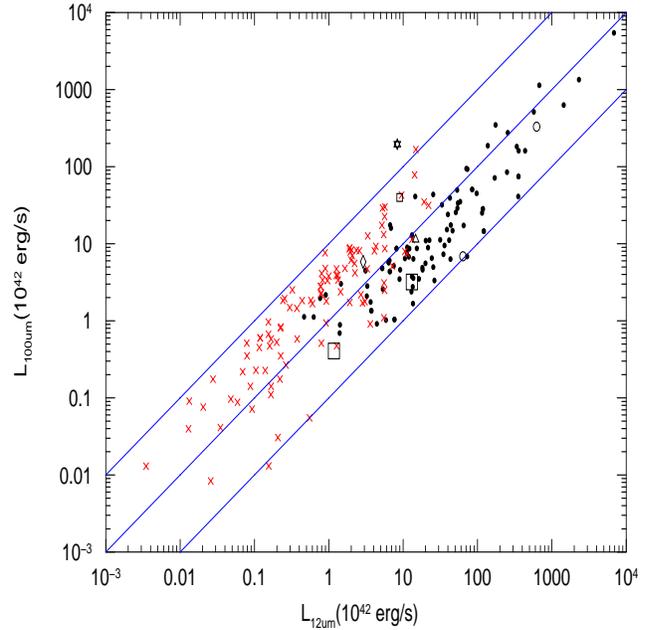}
\caption{Mean 12$\micron$ IR luminosity (from IRAS) plotted against mean 
100$\micron$ IR luminosity for the AGN in our sample. Symbols \& colours are 
as in Fig.~\ref{fig:irx12}. Also plotted are lines of constant luminosity 
ratio ($L_{100\micron}/L_{12\micron}$=10,1,0.1).
\label{fig:ir12v100}}
\end{figure}

Figure~\ref{fig:ir12v100} shows the mean far-IR (100$\micron$) luminosity 
plotted against the mean mid-IR (12$\micron$) luminosity for the 225 AGN 
in our sample without jets, for which both mid-IR and far-IR data were 
available. 115 AGN are group 1 (black circles), 110 are group 2 (red 
crosses). Also plotted are 
lines denoting constant luminosity ratios of $10,1,0.1$ respectively. The
 populations of group 1 and group 2 AGN are very well separated 
by the luminosity ratio 
$L_{100\micron}/L_{12\micron}=1$. Only 22/110 ($\sim 20\%$) of group 2 
have $L_{100\micron}/L_{12\micron}<1$. The three outlier group 2 AGN in this 
plot near the $L_{100\micron}/L_{12\micron}=0.1$ line are nearby LINERs that 
could be cross-classified. From Fig.~\ref{fig:ir12v100} we can say that the 
$L_{100\micron}/L_{12\micron}$ ratios for group 2 AGN are on average an order 
of magnitude larger than those for group 1 AGN. The dispersion in both groups 
is also similar (roughly an order of magnitude). The 
$L_{100\micron}/L_{12\micron}$ ratio also appears to neatly separate out the 
'archetypal' Group 2 AGN, in order of most 'cold' (100$\micron$) dust to least
 'cold' dust. Evidently the starburst Arp 220 (open star) has by far
 the highest value of $L_{100\micron}/L_{12\micron}$ (as we should expect e.g.
 \citet{b25,b29}) followed by the 
Compton-thick AGN seen in transmission (NGC 4945- open square), Circinus 
(open diamond) and NGC 1068 (open triangle) respectively. 

\section{Ruling out selection bias}
\label{sec:blank}
\begin{figure}
\includegraphics[height=3.35in,width=3.35in,angle=-90]{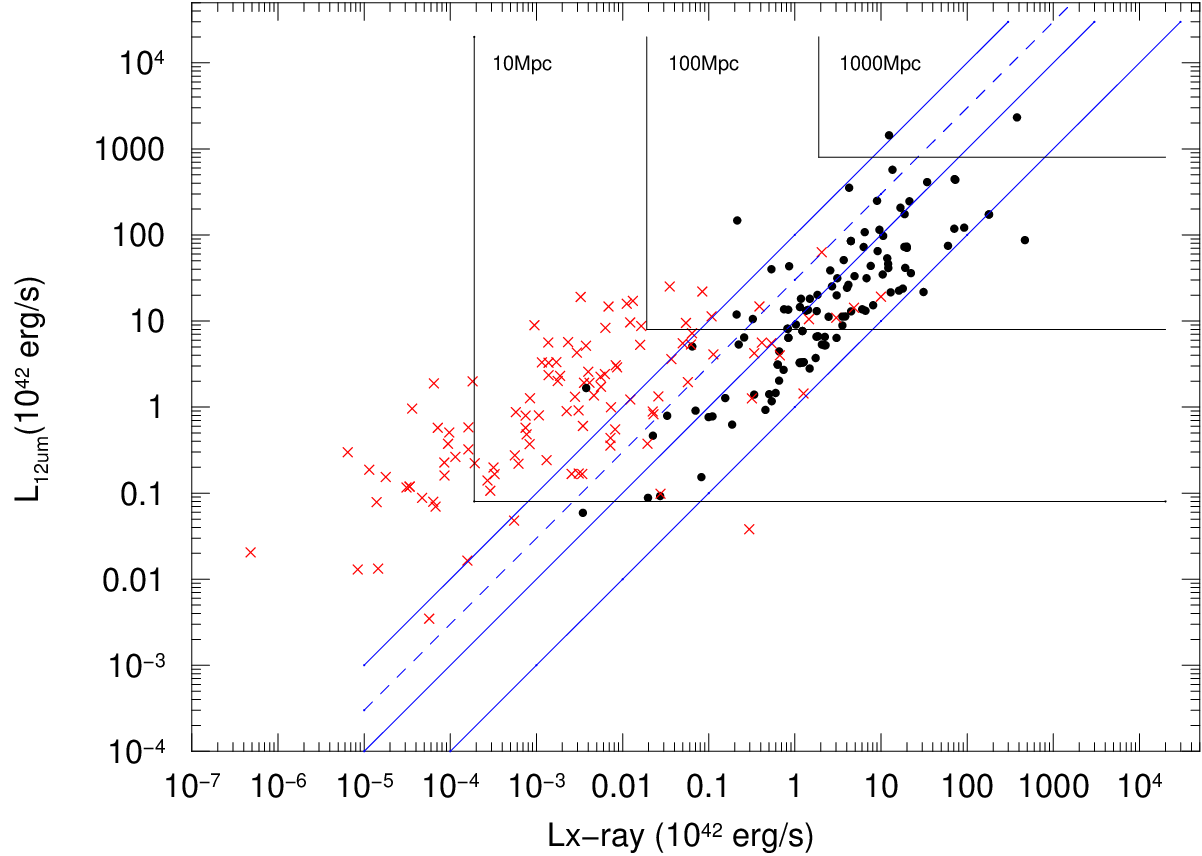}
\caption{As Fig.~\ref{fig:irx12}, except also indicated are the luminosity 
lower limits from the IRAS 12$\micron$ survey (0.3Jy, from \citet{b60}) and 
the ROSAT All-Sky survey ($2\times 10^{-13}$ ergs $\rm{cm}^{-2} \rm{s}^{-1}$ 
from \citet{b95}). The ROSAT limits are from soft (0.1-2.4keV) X-rays and are 
therefore a reasonable proxy flux limit for AGN unlikely to be very heavily 
absorbed in the soft X-ray band (our Group 1 AGN). The IRAS flux limits apply 
to both Group 1 and Group 2 AGN. AGN at $<$10Mpc,$<$100Mpc and $<$1000Mpc 
distances that have not been detected must lie in the region of parameter 
space to the left of the appropriate X-ray limit (Group 1 AGN only) and below 
the appropriate indicated IR limit (both Group 1 and Group 2 AGN).
\label{fig:fluxlimits}}
\end{figure}

Fig.~\ref{fig:irx12} demonstrates that the luminosity ratio 
($R_{ir/x}$) of the 245 AGN in our sample is not a scatterplot. 
So, are the unoccupied regions of the $R_{ir/x}$ plot 
genuinely depopulated, providing a model-independent constraint on AGN 
structure, or is there a selection effect at work? First, flux limits 
from all-sky surveys in both the IR band and the X-ray band allow us to 
establish which regions of $R_{ir/x}$ are probably genuinely unoccupied by AGN.
Fig.~\ref{fig:fluxlimits} is as Fig.~\ref{fig:irx12} except superimposed are 
flux detectability thresholds from (solid line) the ROSAT All-sky survey ($2 
\times 10^{-13}$ ergs $\rm{cm}^{-2} \rm{s}^{-1}$ from \citet{b95}) and the 
IRAS 
12$\micron$ complete sample ($\sim 0.3$Jy from \cite{b60}), corresponding to 
luminosity distances of 10Mpc,100Mpc and 1000Mpc. The ROSAT survey 
was carried out in soft X-rays (0.1-2.4keV) and will undercount heavily 
absorbed AGN (ie Group 2 AGN) in this band. So, since Group 1 AGN are not 
very heavily absorbed in the 
soft X-ray band and since 115/116 of the Group 1 AGN in our sample are 
detectable in the 2-10keV band at ROSAT flux levels, the ROSAT limits are a 
reasonable proxy for X-ray flux limits for Group 1 AGN. This is the best that 
we can do since surveys 
overlapping the 2-10keV band, are neither all-sky nor complete 
\citep{b98,b37}. Distances of a few 
tens of Mpc correspond to a volume of space spanning all varieties of nuclear 
activity, so it seems probable that empty regions in the upper right corner of 
$R_{ir/x}$ are genuinely depopulated by luminous (Group 1) AGN. Any Group 1 AGN
 undiscovered in the X-ray band, must lie to the left of
 their corresponding vertical luminosity distance detection limit (e.g. a 
Type 1 QSO at 100Mpc undiscovered in X-rays, must lie to the left of the 
vertical 100Mpc line). Similarly, any low luminosity AGN, undiscovered at 
12$\micron$ must lie below the corresponding horizontal luminosity distance 
detection limit (e.g. a Type 2 QSO at 100Mpc undiscovered at 12$\micron$ must 
lie below the horizontal 100Mpc line).

\begin{figure}
\includegraphics[height=3.35in,width=3.35in,angle=-90]{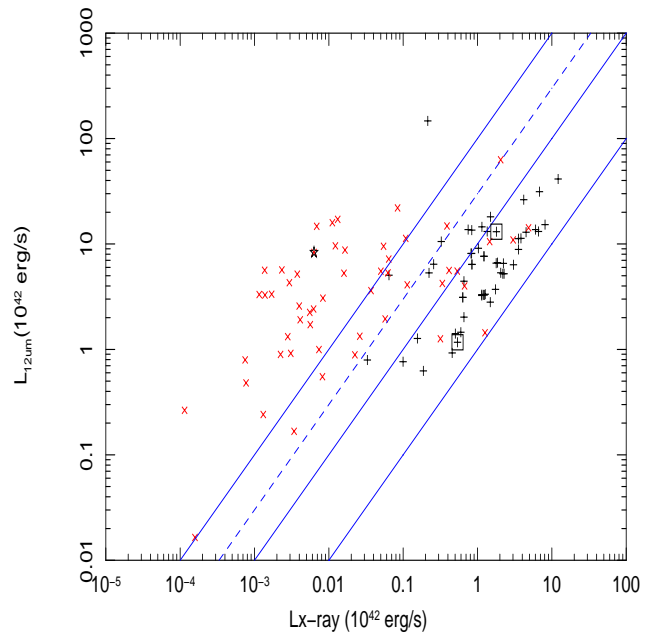}
\caption{As Fig.~\ref{fig:irx12}, except we only plot those AGN in the 
redshift range z=[0.005,0.05] (or $\sim$[20,200]Mpc). For clarity, we denote 
the Group 1 AGN with plus symbols. Note that 53/56 
Group 1 AGN lie in the range $30<R_{ir/x}<1$, a similar ratio to the 103/115 
Group 1 AGN in Fig.~\ref{fig:irx12}.  
\label{fig:cz}}
\end{figure}

Second, we also need to check that the dispersion of AGN is not unduly 
affected by picking up higher luminosity AGN at relatively higher redshifts. 
In Fig.~\ref{fig:cz} we plot the 109 AGN (56 Group 1 and 53 Group 2 AGN) in 
the redshift range $z=[0.005,0.05]$ (or $\sim$[20,200]Mpc). Fig.~\ref{fig:cz} 
excludes the highest and lowest luminosity AGN from both Groups. From 
Fig.~\ref{fig:cz}, 53/56 Group 1 AGN have $1<R_{ir/x}<30$, a similar ratio to 
the 103/115 Group 1 AGN in the same range, using our full sample. The Spearman 
rank correlation coefficient ($S_{rc}$) for our complete Group 1 distribution 
(without jets) is 0.69 with a strong statistial significance 
($5 \times 10^{-8}$). When we split our
 Group 1 distribution into those AGN at $z<0.05$ (65/112) and $z>0.05$ 
(47/112), we find that $S_{rc}$ is statistically significant at $0.71$ and 
$0.67$ respectively. We also find that the mean value of $R_{ir/x}$ for 
both these $z<0.05$ and $z>0.05$ Group 1 AGN populations are consistent with 
each other and with the mean value of $R_{ir/x}$ of the complete Group 1 
population, within the mean absolute deviation of each population. 
Furthermore, the T-statistic for these 'low' and 'high' z 
Group 1 AGN has a significance of 0.5, indicating that the $z<0.05$ and 
$z>0.05$ populations do not originate in distributions with substantially 
different mean values of $R_{ir/x}$. In future work, we anticipate extending 
our sample with a view to obtaining strong statistical constraints on the value of $R_{ir/x}$ in AGN.

\begin{figure}
\includegraphics[height=3.35in,width=3.35in,angle=-90]{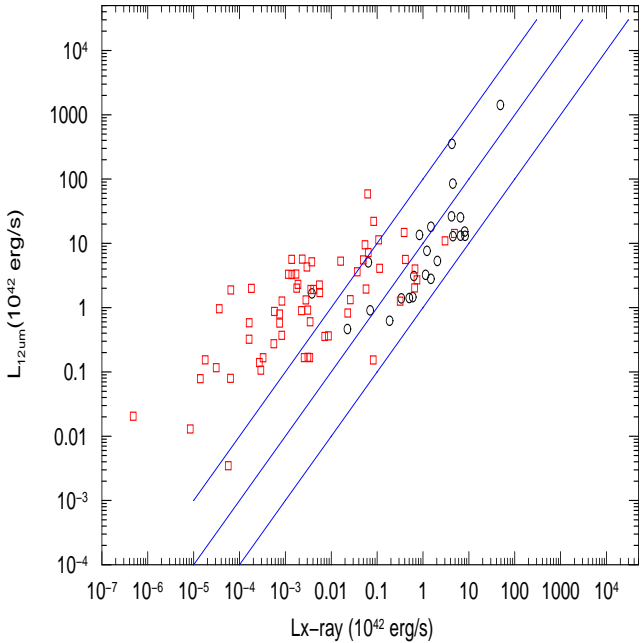}
\caption{As Fig.~\ref{fig:nojets}, except we only plot those AGN from 
the complete 12$\micron$ IRAS flux-limited sample \citet{b59} with 2-10keV 
X-ray data. Open red squares denote Group 2 AGN and open black 
circles denote Group 1 AGN from \citet{b59}.
\label{fig:iras12sample}}
\end{figure}

Another way of assessing the depopulation of regions of the $R_{ir/x}$ plot 
is by considering the dispersion of complete samples of AGN. The samples of 
\citet{b59} and \citet{b60} in Table~\ref{tab:sample} 
correspond to complete, flux-limited IRAS samples at 12$\micron$ and 
$60\micron$ respectively. However, in each case, only $\sim 15\%$ of each 
sample have 2-10keV X-ray data. However, the sample of \citet{b59} in 
particular spans most of our range of $L_{x},L_{12um}$ space, so a comparison 
of our sample with \citet{b59} could tell 
us if we are under-sampling a particular region of parameter space. 
Fig.~\ref{fig:iras12sample} is as Fig.~\ref{fig:nojets}, except that only AGN from \citet{b59} are plotted.

\begin{figure}
\includegraphics[height=3.35in,width=3.35in,angle=-90]{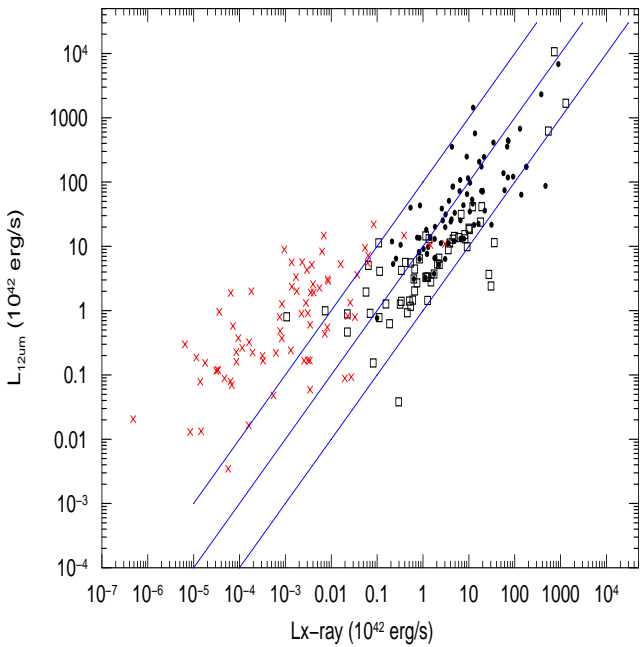}
\caption{As Fig.~\ref{fig:nojets}, except also indicated  as open circles are 
those AGN in the BAT X-ray sample \citet{b37}, which is unlimited by X-ray 
absorption.
\label{fig:bat}}
\end{figure}

Clearly, the subset of the complete 12$\micron$ IRAS AGN sample span the full 
range of $R_{ir/x}$ for Group 2 AGN and most of the range of $R_{ir/x}$ for 
Group 1 AGN. Furthermore, the 12$\micron$ sample subset does not stake out 
otherwise 
unoccupied regions of $R_{ir/x}$ space. It is possible that the remainder of 
the complete IRAS samples picked out AGN that live in e.g. the top-left of 
Fig.~\ref{fig:nojets}, i.e. low $L_{x}$ and large $L_{ir}$. However, an 
additional $\sim 20\%$ IRAS AGN had $0.2-4.0$keV \emph{Einstein} X-ray flux 
measurements, yielding similar values of $R_{ir/x}$ to those in 
Fig.~\ref{fig:iras12sample} (see also 
\citet{b62} and \citet{b69}). The sample of AGN studied by \citet{b37} is not 
limited by X-ray absorption, 
so it is useful to consider the range of $R_{ir/x}$ spanned by these AGN. 
Fig.~\ref{fig:bat} is as Fig.~\ref{fig:nojets}, but indicates those AGN 
from \citet{b37}. Once again, it is clear that this sample spans the existing 
ranges of $R_{ir/x}$ space and does not stake out otherwise unoccupied regions.

In order to test the effect of AGN redshift on our sample, we studied an 
independent sample of 150 additional high-z AGN observed from deep surveys with XMM-Newton and Chandra \citep{b28,b27,b32}. At the high 
redshifts of AGN in the deep surveys (typically z$\sim 1$), an observed IR 
luminosity at $24\micron$ corresponds \emph{approximately} to an AGN-frame 
$12\micron$ IR luminosity (see e.g. Treister, Krolik \& Dullemond (2008)) and 
a 2-10 keV X-ray luminosity corresponds to an 
AGN-frame $4-20$ keV X-ray luminosity (which will typically lie within a 
factor of a few of the actual 2-10keV luminosity), so luminosity ratios 
$R_{ir/x}$ for these deep AGN will systematically be a factor of a few higher 
on average than for the local AGN in Fig.~\ref{fig:irx12}.

\begin{figure}
\includegraphics[height=3.35in,width=3.35in,angle=-90]{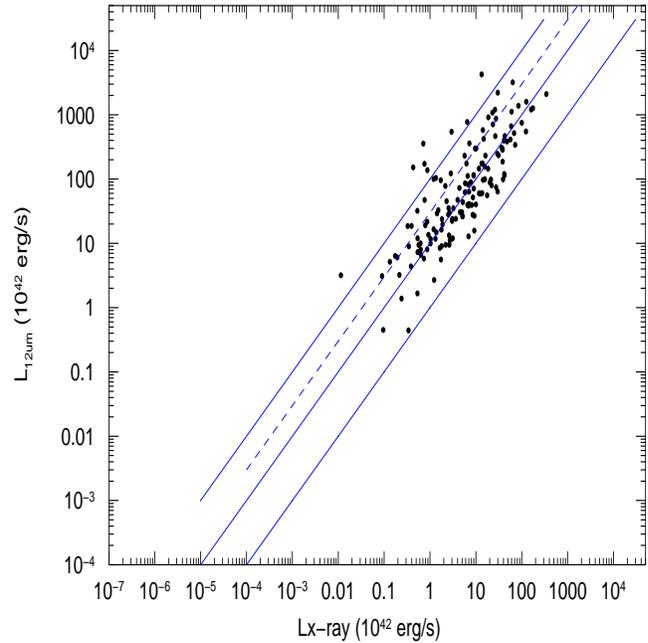}
\caption{As Fig.~\ref{fig:irx12}, except we plot filled-in circles a new 
sample of 150 AGN from the deep surveys of \citet{b28,b27,b32}. The observed 
luminosity ratio 
for these deep ($z\sim 1$) AGN uses 2-10keV observed fluxes which correspond 
to $\sim 4-20$keV in the AGN restframe. Therefore the luminosity ratio 
$R_{ir/x}$ will be a factor of a few higher (ie. displaced slightly leftwards 
on the plot) on average than the Group 1 AGN in Fig.~\ref{fig:irx12}.
\label{fig:deep}}
\end{figure}

From Fig.~\ref{fig:deep}, even without considering the systematic 
overestimate of $R_{ir/x}$ above, the high z AGN appear to closely follow 
the luminosity ratio 
for the Group 1 AGN in Fig.~\ref{fig:irx12} and very
 few lie in the Group 2 AGN region (8/150 have $R_{ir/x}>100$) or in the blank
 areas of our luminosity plot. While there are clearly 
many caveats to using such deep survey data (not least assumptions about the 
AGN SED shape based on the IR spectrum), it is intruiging that, even with our 
very simple approach, the deep survey points should follow very closely 
the low-z Group 1 AGN dispersion and not fill blank regions of the luminosity 
ratio plots. It is also comforting that our main distribution of Group 1 AGN 
in Fig.~\ref{fig:irx12} is also consistent with the dispersion of the 
population of AGN in \citet{b71}, which, since the X-rays are 
14-195keV, should be independent of X-ray absorption.

\section{Discussion}
\label{sec:discuss}
Much of the IR emission from active galactic nuclei should come from 
reprocessed X-ray emission. There should also be some component of intrinsic 
IR emission in the SED (including non-thermal IR; although we shall ignore the
 complication of jets in the following discussion) and any remaining nuclear 
IR emission should be due to star formation in the thick dust in the 
nucleus. Whether the dust lives in a torus (doughnut) or in clumps, we can 
characterize the IR emission from 'active' nuclei (without prominent jets) as:
\begin{equation}
\begin{tabular}{lcl}
$\rm{L}_{IR}$&=&$\rm{L}_{AGN} + \rm{L}_{stars}$\\
&=&$\rm{L_{x}} g_{unit}\left[g_{sed} + \left(\frac{\Omega}{4\pi}\right) f_{abs} g_{repro} 
f_{Lbol} \right] + \rm{L}_{stars}$ 
\end{tabular}
\label{eq:one}
\end{equation}
where $\rm{g_{unit}}$ is a numerical factor from converting $\rm{L_{x}}$, 
the 2-10keV intrinsic X-ray luminosity to the same units as $\rm{L_{IR}}$, 
$\rm{g_{sed}}$ is a factor corresponding to intrinsic IR emitted by the AGN 
(normalized to $\rm{L_{x}}$), $\Omega/4\pi$ is the covering factor of the 
reprocessing material, $\rm{f_{abs}}$ is the 
fraction of the 2-10keV radiation absorbed, $\rm{g_{repro}}$ is 
a factor depending on the details of the reprocessing physics (including 
dust size \& chemistry), $\rm{f_{Lbol}}$ is the 
fraction of the bolometric luminosity absorbed by the cool dust.

For most group 1 AGN we expect, $\rm{L}_{AGN}\gg \rm{L}_{stars}$, so 
the observed luminosity ratio for most group 1 AGN should be
\begin{equation}
\frac{\rm{L}_{IR}}{\rm{L_{x}}} \approx \rm{R_{i/o}g_{unit} [g_{sed} +
\left(\frac{\Omega}{4\pi}\right)f_{abs} g_{repro} f_{Lbol}]}
\label{eq:two}
\end{equation}
where $\rm{R_{i/o}}$ is the ratio of intrinsic to observed 2-10keV X-ray 
luminosity and is a function of the column of X-ray absorbing material along 
the observers' sightline. From Figs.~\ref{fig:irx12} and \ref{fig:irx100}, the
 dispersion 
in $R_{ir/x}$ for most ($\sim $90\%) group 1 AGN is a factor of 
30 or less. Therefore the total combined variation in all the multipliers in 
equation~\ref{eq:two} must be less than thirty for 
most Group 1 AGN. The factor $\rm{f_{abs}}$ is a function both of the covering 
fraction of the reprocessing material and its column density 
($\rm{f_{abs}=f_{c}f_{Nh}}$). Using XSPEC, we 
tested the effect of column density on $\rm{f_{abs}}$ by investigating
 the change in X-ray energy for absorbing 
column density in the range $N_{H}=[10^{19},10^{24}] \rm{cm}^{-2}$. We found 
that a factor of $10^{5}$ increase in absorbing column density, only produced 
a factor of five in the X-ray energy absorbed (similar to the 
results of \citet{b1} and \citet{b25}). So a 'naked' Group 1 AGN without 
substantial amounts of absorbing material (e.g. a torus) in its environment, 
should have an IR flux only $\sim 1/5$th that of a Group 1 AGN surrounded by 
substantial amounts of dust (not necessarily in the sightline).  Therefore, a 
\emph{total} dispersion 
of $1<R_{ir/x}<30$ for most Group 1 AGN, will include a factor of $\leq 5$ for 
$f_{Nh}$, indicating that $\rm{f_{unit},g_{sed}, g_{repro},f_{Lbol}, f_{c}}$ 
could vary collectively by 
as little as a factor of six, across all Group 1 AGN.  Thus, the covering 
factor and fraction of absorbing material (clouds, torus, flared disk etc), 
the composition of the material (dust chemistry, abundance, size), the column 
density of the material and the SED X-ray to IR ratio in group 1 AGN 
\emph{collectively} vary by a factor of thirty or less and possibly by as 
little as a factor of six. Note that this is a 
\emph{model-independent result}. We have not assumed that the absorbers must 
either be in a torus or clouds or some combination. Such a tight dispersion 
among Group 1 AGN severely restricts models of accretion, geometry 
and physical conditions in AGN. These Group 1 AGN 
collectively span a range of $\sim 10^{3}$ in central black hole 
mass (e.g. \citet{b58}), suggesting that independent of jets, the central 
engines of Group 1 AGN are quite \emph{remarkably} similar. Again, it is worth 
noting that this remarkably 
small variation in AGN properties is based on \emph{observed} quantities and is
therefore model-independent.
 
Group 2 AGN, with associated 
star-forming regions, could have $\rm{L}_{AGN} < \rm{L}_{stars}$ in 
equation~\ref{eq:one}. Furthermore, Fig.~\ref{fig:ir12v100} indicates 
that the ratio of cold dust to warm dust (indicative of star formation) is an 
order of magnitude larger in group 2 AGN than in group 1 AGN. The fact 
that dispersion in $R_{ir/x}$ is 2-3 orders of 
magnitude larger for Group 2 AGN than for Group 1 AGN, indicates either that 
(a) observed $\rm{L_{x}}$ $\ll$ intrinsic $\rm{L_{x}}$ or that (b) $L_{stars}$
 $\gg$ $L_{AGN}$ in Group 2 AGN or both. \citet{b71} find that the intrinsic 
X-ray luminosity for Group 2 AGN is comparable to that for Group 1 AGN, so in 
most cases observed $\rm{L_{x}}$ is probably substantially lower than 
intrinsic $\rm{L_{x}}$. It is also possible that $\rm{L_{stars}}$ is 
intrinsically greater in group 2 AGN than in group 1 AGN.

\section{Conclusions}
\label{sec:conclusions}
Our aim was to compare common observed AGN properties that encompass the 
continuum of AGN activity in an effort to, if possible, derive 
model-independent constraints on AGN structure and the accretion 
neighbourhood. We studied the observed X-ray to IR luminosity 
ratios of a heterogeneous sample of 245 galaxies from the literature and, in 
spite of our simple approach, AGN variability and non-simultaneity of 
most of the data, we found strong trends.  

Fig.~\ref{fig:irx12} reveals the emergence of observed 'activity' at 
$L_{IR}, L_{X} >10^{42} \rm{erg} \rm{s}^{-1}$. Below these luminosities, AGN 
may compete with X-ray and IR emitters in the host galaxy (such as ULXs, see 
e.g. \citet{b30}) or may be obscured 
by large columns of absorbing material. Since emission from jets considerably 
complicates our interpretation of the observed luminosity ratios, we only 
considered those AGN without prominent observed jets. We find that AGN 
considered 'face-on' or unobscured in the standard model (our group 1) are 
tightly clustered by observed luminosity ratio ($\sim 90\%$ have 
$1<R_{ir/x}<30$), indicating 
that there is very little variation between individual group 1 AGN in: 
(a) intrinsic SED (i.e. accretion flow, whether ADAF, tubulent disk or both), 
(b) solid angle of the reprocessing ($X \rightarrow IR$) absorber (whether 
clouds, torus, flared-disk or some combination) and (c) details of the 
absorbing material (chemistry \& dust grain sizes). Thus, 
Fig.~\ref{fig:irx12} 
demonstrates that underlying Group 1 AGN is a remarkably uniform phenomenon. 

Fig.~\ref{fig:irx12} shows a clear difference in $R_{ir/x}$ between 
unobscured AGN (group 1) and obscured \& low luminosity AGN (our group 2). 
Despite the fact that our sample is \emph{not} complete, our findings 
for the dispersion of AGN are confirmed in a comparison with other, 
independent AGN samples (e.g. \citet{b37,b71}, which are not limited by X-ray 
absorption, or \citet{b32,b27,b28} from $z\sim 1$). Furthermore, subsets of our
 sample of 245 AGN are derived from complete IRAS samples \citep{b60,b59} and 
these subsets at least do not span values of $R_{ir/x}$ different from the 
rest of 
our sample. Bias is of course a potential problem with any approach such as 
ours, but it is comforting that we can find some independent verification for 
our simple, model-independent approach. Estimates of the intrinsic 2-10
 keV luminosity for Compton-thick AGN from very hard X-ray (absorption 
independent) observations 
(e.g. \citet{b37,b38,b36}) reveal that the intrinsic luminosity ratio 
is consistent with the group 1 AGN (in agreement with the findings of e.g. 
\citet{b3,b4,b5,b71}). However, unlike \citet{b3,b4,b5} we have endeavoured to
 avoid the highly model-dependent business of estimating intrinsic 2-10keV 
X-ray luminosity of group 2 AGN based on the 2-10keV observed luminosity. We 
also studied the far-IR to mid-IR luminosity 
ratios of the nuclei in our 
sample. We found that the $100\micron$ to $12\micron$ luminosity ratio 
($L_{100\micron}/L_{12\micron} \sim 1$) is an 
extremely effective separator of the group 1(unobscured) and group 2 
(obscured \& low luminosity) populations (see Fig.\ref{fig:ir12v100}). 
Using archetypal AGN as a guide, this plot ranks group 2 
AGN in order of decreasing luminosity ratio for increasing obscuration. 

In principle our approach could be applied to intermediate mass black holes 
(IMBHs) and Galactic black hole candidates (GBHCs) to investigate the 
accretion neighbourhood around black holes on all mass scales. The outstanding
 question is whether reprocessing material around IMBHs and 
GBHCs is substantially different in covering factor, structure, chemistry or 
reprocessing details from that around AGN. Are a flared disk or dusty torus 
associated with a minimum mass scale, or are we looking at a basically 
identical phenomenon across 7-8 orders of magnitude in mass?

\section*{Acknowledgements}
We gratefully acknowledge support from NASA grant GO6-7085B(BM). We made 
extensive use of the NASA/IPAC 
Extragalactic Database (NED), operated by the Jet Propulsion Laboratory, 
CalTech, under contract with NASA. We gratefully acknowledge very helpful 
discussions with Tahir Yaqoob who helped inspire this work and with Sylvain 
Veilleux who provided useful insights into heterogeneous samples and samples 
of IR-observed AGN. BM \& KESF gratefully acknowledge the 
support of the Department of Astrophysics of the American Museum of Natural 
History and PSC-CUNY grants PSCOOC-38-99 and PSCOOC-38-98 respectively.

%\bsp

\label{lastpage}

\end{document}